\documentclass[aps, reprint,prl,nofootinbib]{revtex4-1}
\usepackage{dcolumn}
\usepackage[utf8]{inputenc}
\usepackage{amsmath}
\usepackage{amsfonts}
\usepackage{amssymb}
\usepackage{graphicx}
\usepackage{hyperref}
\usepackage{xcolor}

\newcommand{\mb}{\mathbf}
\newcommand{\bs}{\boldsymbol}

\newcommand{\rme}{\mathrm{e}}

\begin{document}

\title{Aharonov-Casher and shielded Aharonov-Bohm effects with a quantum electromagnetic field}

\author{Pablo L. Saldanha}\email{saldanha@fisica.ufmg.br}
\affiliation{Departamento de F\'isica, Universidade Federal de Minas Gerais, Belo Horizonte, MG 31270-901, Brazil}

\date{\today}

\begin{abstract}
{We use a covariant formalism that is capable of describing the electric and magnetic versions of the Aharonov-Bohm effect, as well as the Aharonov-Casher effect, through local interactions of charges and currents with the quantum electromagnetic field. By considering that only local interactions of a quantum particle with the quantum field can affect its behavior, we show that the magnetic Aharonov-Bohm effect must be present even if the solenoid generating the magnetic field is shielded by a perfect conductor, as experimentally demonstrated. 
}
\end{abstract}


\maketitle

\section{Introduction}

The Aharonov-Bohm (AB) effect \cite{ehrenberg49,aharonov59} is a striking feature of quantum mechanics that completely changed our notions of electromagnetic interactions compared to the classical view. Classical electrodynamics can be understood in terms of local interactions between charges and the electric and magnetic fields. The AB effect, on the other hand, is usually described by a local interaction of a quantum charged particle with the electromagnetic potentials \cite{aharonov59} or by a nonlocal influence of electromagnetic fields, charges or currents on the particle \cite{liebowitz65,boyer73,boyer02,peshkin81,vaidman12,kang13,kang15,saldanha16,pearl17a}. 
This is because in the magnetic AB effect, for instance, a quantum particle is subjected to a null electromagnetic field (but a nonzero vector potential) in the two possible paths of an interferometer that encloses a long solenoid, but its interference pattern depends on the presence of fields and currents in the solenoid. 

In order to solve the conceptual issues regarding the locality of the AB effect, previous works have treated this phenomenon considering a quantized electromagnetic field \cite{santos99,choi04,pearl17b,li18,marletto19,saldanha19}. In this way, the interaction between the quantum particle and the solenoid can be understood as an exchange of photons between these entities, where both the particle and the solenoid interact locally with the quantum electromagnetic field. However, these previous works do not explain how the AB effect can be described when the solenoid is covered by a perfect conductor shield \cite{aharonov15}. 
In this case, one might expect that a photon exchange between the particle and the solenoid would not be possible. Since the AB effect with shielding was experimentally observed \cite{tonomura86}, this is an important issue to address.

Here we show how the AB effect with a perfect conductor shield involving the solenoid can also be explained in terms of local interactions, with the use of a quantum electromagnetic field. To do that we consider three current densities, associated to the quantum particle, the solenoid, and the shield, that interact locally with the quantum electromagnetic field. We present our results in a unified covariant formalism that is capable of predicting the electric and magnetic versions of the AB effect \cite{aharonov59} (with and without shielding), as well as the Aharonov-Casher (AC) effect \cite{aharonov84}, which, to our knowledge, wasn't treated with a quantum field yet. 

 \section{A unified quantum model} 

We will describe all the AB effects and the AC effect depicted in Fig. 1 with a quantum treatment for the electromagnetic field. The different subsystems interact locally with the quantum electromagnetic field, that mediates the interactions between them. Each subsystem (quantum particle, solenoid, conductor shield, etc.) will be treated as a charge density $\rho_i(\mb{r}_i)$ and a current density $\mb{J}_i(\mb{r}_i)$. The index $i=1$ will be reserved for the quantum particle, and when it is a charged particle (spinless for simplicity) we have
\begin{equation}\label{J1}
	\rho_1(\mb{r}_1)=q|\Psi_1(\mb{r}_1)|^2,\;\;\mb{J}_1(\mb{r}_1)=\frac{q}{2m}[\Psi_1^*\mb{p}_1\Psi_1-\Psi_1\mb{p}_1\Psi_1^*],
\end{equation}
where $q$ and $m$ are the particle charge and mass, $\Psi_1$ its wavefunction and $\mb{p}_1$ the momentum operator. We consider nonrelativistic quantum particles in this work. In the AC effect the quantum particle is a neutral particle with spin, and we treat this case latter. The other charge and current densities in the schemes are to be considered classical.

\begin{figure}
  \centering
    \includegraphics[width=8.5cm]{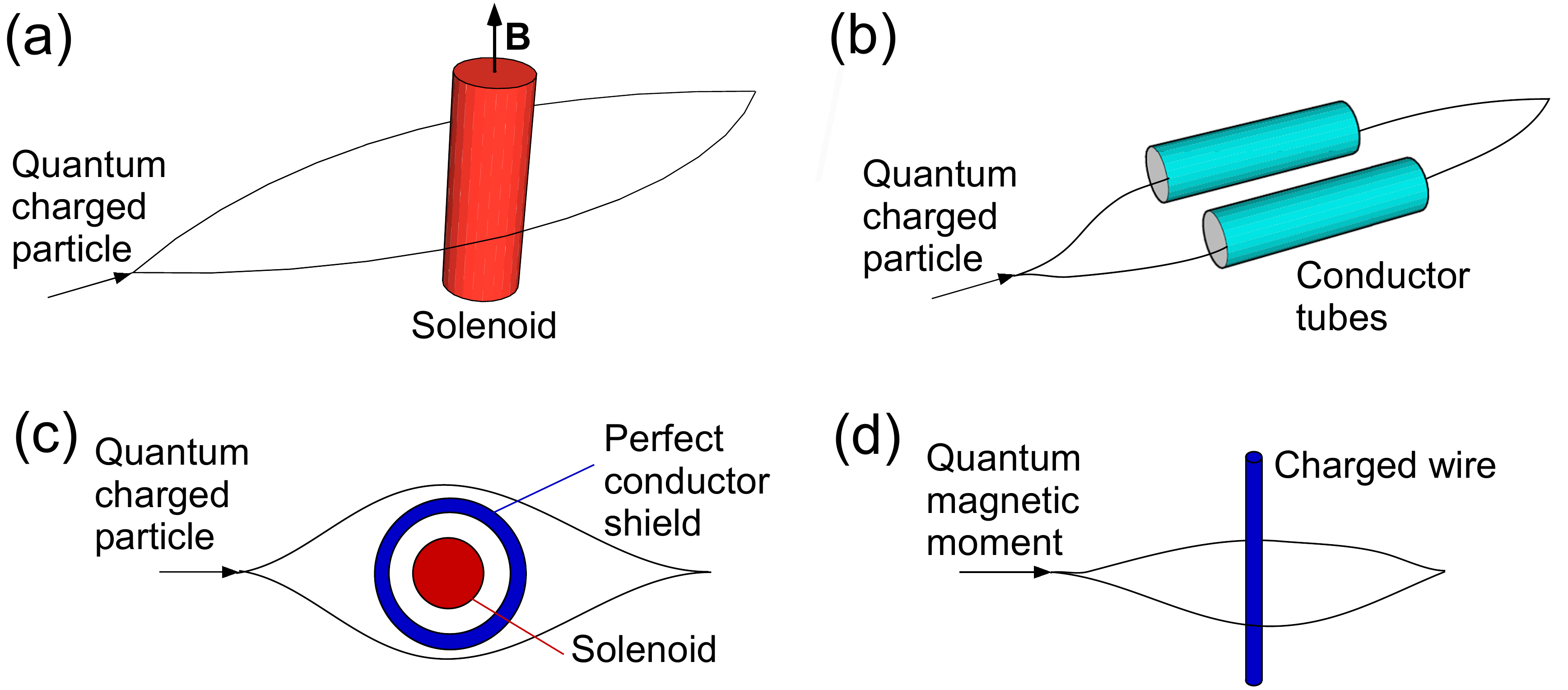}
  \caption{Aharonov-Bohm (AB) and Aharanov-Casher (AC) schemes. (a) Magnetic AB effect. An interferometer encloses a long solenoid. The interference pattern of quantum charged particles depend on the magnetic flux inside the solenoid even if they only propagate in regions where the magnetic field is zero. (b) Electric AB effect. The electrostatic potentials of the tubes are different from zero only when a charged particle is in a superposition of being inside each tube. The interference pattern depends on these potentials even if the particles only propagate in regions where the electric field is zero. (c) Magnetic AB effect with a perfect conductor shield involving the solenoid. (d) AC effect. An interferometer encloses a charged wire. The interference pattern of neutral particles with magnetic moment depend on the wire charge even if there is no force on these particles.}
\end{figure}

The interaction Hamiltonian between the different charge and current densities and the quantum electromagnetic field, considered to be not very intense,  can be written as
\begin{equation}\label{v1}
	V=\sum_i V_i, \;\;\mathrm{with}\;\; V_i=\int d^3r_i\left[\rho_iU(\mb{r}_i) -\mb{J}_i\cdot \mb{A}(\mb{r}_i)\right],
\end{equation}
where 
\begin{equation}\label{A}
	\mb{A}(\mb{r})=\int d^3k \sum_\sigma \sqrt{\frac{\hbar}{2\varepsilon_0\omega(2\pi)^3}}\;\hat{\epsilon}_{\mb{k}\sigma}a_\sigma(\mb{k})\rme^{i\mb{k}\cdot\mb{r}}+\mathrm{H.c.}
\end{equation}
is the potential vector operator, where $\hat{\epsilon}_{\mb{k}\sigma}$ is a polarization vector, $a_\sigma(\mb{k})$ is the annihilation operator for a mode with wavevector $\mb{k}$ and polarization index $\sigma$, and $\omega=ck$ is the mode angular frequency, and
\begin{equation}\label{U}
	U(\mb{r})=c\int d^3k \sqrt{\frac{\hbar}{2\varepsilon_0\omega(2\pi)^3}}a_0(\mb{k})\rme^{i\mb{k}\cdot\mb{r}}-\mathrm{H.c.}
\end{equation}
is the scalar potential operator \cite{cohen}.  We are using the Lorenz gauge for the fields, such that the index $\sigma$ in Eq. (\ref{A}) assumes 3 values for each wave vector $\mb{k}$, with the presence of longitudinal photons. $a_0(\mb{k})$ in Eq. (\ref{U}) is an annihilation operator for a scalar photon with wave vector $\mb{k}$. So this treatment differs from the usual quantum optics formalism in the Coulomb gauge, where only photons with transverse polarizations are considered \cite{mandel}. The consideration of 4 kinds of photons for each wave vector $\mb{k}$ is essential for constructing a covariant theory. It can be shown that longitudinal and scalar photons cannot carry energy or momentum between different regions of space \cite{cohen}, such that they play no role in the exchange of energy between the electromagnetic field and matter or in light propagation, which are the topics usually treated in quantum optics textbooks \cite{mandel}. For this reason the Coulomb gauge is simpler to use in these cases. However, longitudinal and scalar photons play important roles in some electromagnetic interactions and must be kept in the present treatment. For instance, the Coulomb interaction between two electric charges can be associated to an exchange of scalar photons between them \cite{cohen}. Note also the minus sign present in the Hermitian conjugate term in Eq. (\ref{U}), which is important for the consistency of the quantum electrodynamics theory \cite{cohen}.

The term $V$ from Eq. (\ref{v1}) in the system Hamiltonian changes the energy of the unperturbed electromagnetic vacuum state $|\mathrm{vac}\rangle$. Since all terms of $V$ have annihilation or creation operators due to the presence of the operators $\mb{A}$ and $U$ from Eqs. (\ref{A}) and (\ref{U}), the first-order correction of the electromagnetic vacuum energy is zero. The second-order correction can be written as \cite{cohen2}
\begin{equation}
	\Delta E= \sum_{n\neq0}\sum_m\frac{|\langle \phi_n^m|V|\mathrm{vac}\rangle|^2}{E_0^0-E_n^0},
\end{equation}
where $E_0^0$ is the energy of the unperturbed vacuum, $|\phi_n^m\rangle$ represent eigenstates of the unperturbed Hamiltonian with energies $E_n^0$, and the index $m$ is associated to the degeneracy of these eigenstates (the vacuum state is nondegenerate). Disregarding self-energy terms, the correction in energy of the electromagnetic vacuum state due to the presence of the electric charges and currents densities can be written as $\sum_{i,j>i}\Delta E_{ij}$, with
\begin{equation}\label{deltaE}
\Delta E_{ij}=2\mathrm{Re}\Bigg[\int d^3k\sum_{\sigma'}\frac{\langle\mathrm{vac}|V_j|\mb{k},\sigma'\rangle\langle \mb{k},\sigma'|V_i|\mathrm{vac}\rangle}{-\hbar\omega}\Bigg],
\end{equation}
where $|\mb{k},\sigma'\rangle$ represents a single-photon state, $\mb{k}$ being its wavevector, $\sigma'$ the polarization index (including $\sigma'=0$ for scalar photons), and $\hbar\omega$ the photon energy. Note that the term explicitly written in the above equation can be associated to the creation of a photon $|\mb{k},\sigma'\rangle$ in subsystem $i$ by the action of the operator $V_i$ in the electromagnetic vacuum, which is destroyed in subsystem $j$ by the action of the operator $V_j$. The Hermitian conjugate represents an inverse photon exchange. So the modification of the electromagnetic vacuum energy in $\Delta E_{ij}$ can be associated to an exchange of photons between these two subsystems. 

Using Eqs. (\ref{v1}), (\ref{A}), and (\ref{U}) we obtain
\begin{eqnarray}\nonumber
	\langle \mb{k},\sigma|V_i|\mathrm{vac}\rangle &=& -\int d^3r_i \;\sqrt{\frac{\hbar}{2\varepsilon_0\omega(2\pi)^3}}\;\mb{J}_i(\mb{r}_i)\cdot\hat{\epsilon}_{\mb{k}\sigma}\;\rme^{-i\mb{k}\cdot\mb{r}_i},\\
	\langle \mb{k},0|V_i|\mathrm{vac}\rangle &=& -\int d^3r_i \;\sqrt{\frac{\hbar}{2\varepsilon_0\omega(2\pi)^3}}\;c\rho_i(\mb{r}_i)\;\rme^{-i\mb{k}\cdot\mb{r}_i},
\end{eqnarray}
such that Eq. (\ref{deltaE}) becomes
\begin{eqnarray}\label{deltaEl}\nonumber
	\Delta E_{ij}&=&\mu_0\mathrm{Re}\Bigg\{\int d^3r_i\int d^3r_j\Bigg[\int d^3k \frac{\rme^{i\mb{k}\cdot(\mb{r}_i-\mb{r}_j)}}{(2\pi)^3k^2}\Bigg]\times\\
	&&\;\;\;\;\;\;\;\;\;\times\left[c^2\rho_i(\mb{r}_i)\rho_j(\mb{r}_j)-\mb{J}_i(\mb{r}_i)\cdot\mb{J}_j(\mb{r}_j)\right]\Bigg\}.
\end{eqnarray}
The term inside brackets in Eq. (\ref{deltaEl}) can be written as 
\begin{equation}
	\int d^3k \frac{\rme^{i\mb{k}\cdot(\mb{r}_i-\mb{r}_j)}}{(2\pi)^3k^2} = \frac{1}{4\pi|\mb{r}_i-\mb{r}_j|}.
\end{equation}
So we can write
\begin{equation}\label{deltaEE}
	\Delta E_{ij}=\int d^3r_i \int d^3r_j \frac{[\rho_i(\mb{r}_i)\rho_j(\mb{r}_j)/\varepsilon_0-\mu_0\mb{J}_i(\mb{r}_i)\cdot\mb{J}_j(\mb{r}_j)]}{4\pi|\mb{r}_i-\mb{r}_j|}.
\end{equation}

Note that if we define the following effective scalar and vector potentials in the Lorenz gauge due to the charge and current densities $\rho_j$ and $\mb{J}_j$:
\begin{equation}\label{aa}
	{\mathcal{U}}_j(\mb{r})=\int d^3r_j \frac{\rho_j(\mb{r}_j)}{4\pi\varepsilon_0|\mb{r}-\mb{r}_j|}, \;\mb{\mathcal{A}}_j(\mb{r})=\int d^3r_j \frac{\mu_0\mb{J}_j(\mb{r}_j)}{4\pi|\mb{r}-\mb{r}_j|},
\end{equation}
we can write
\begin{equation}\label{DE}
	\Delta E_{ij}=\int d^3r_i [\rho_i(\mb{r}_i)\mathcal{U}_j(\mb{r}_i)-\mb{J}_i(\mb{r}_i)\cdot\mb{\mathcal{A}}_j(\mb{r}_i)].
\end{equation}
Note that we can choose which subsystem acts as a source of the effective potentials and which subsystem acts as interacting with these potentials in the above equation, since $\Delta E_{ij}=\Delta E_{ji}$. For instance, in the scheme of Fig. 1(a) we can consider the effective vector potential of the solenoid acting on the quantum charge or the effective potentials of the charge acting on the solenoid current (similar to Vaidman's treatment \cite{vaidman12}, although only fields, not potentials, are considered there). It is important to stress that the effective potentials are to be seen only as mathematical tools to write the change of the electromagnetic vacuum energy in a simple way in Eq. (\ref{DE}), not as physical quantities.

\section{Aharonov-Bohm effect} 

Now let us proceed to apply the formalism developed so far to the different AB schemes presented in Fig. 1. In Fig. 1(a) we have the magnetic version of the AB effect, where a quantum charged particle has two possible paths in an interferometer that encloses a long solenoid. Consider $\mb{\mathcal{A}}_2$ as the effective potential vector generated by the current density $\mb{J}_2$ of the solenoid and $\mb{J}_1$ the quantum particle current density given by Eq. (\ref{J1}). Writing the particle wave function as a superposition of wave packets that have a reasonably well defined momentum $\mb{p}_{n}(t)$ and position $\mb{r}_{n}(t)$ along their propagation, the accumulated AB phase for each wave packet can be written as 
\begin{equation}
	-\int dt \frac{\Delta E_{12}}{\hbar}=\frac{q}{\hbar}\int dt \,\mb{v}_{n}\cdot \mb{\mathcal{A}}_2(\mb{r}_{n}) =\frac{q}{\hbar}\int \mb{dl}_{n}\cdot{\mathcal{A}}_2,
\end{equation}
where $\Delta E_{12}$ is given by Eq. (\ref{DE}), $\mb{v}_{n}\equiv\mb{p}_{n}(t)/m$ is the wave packet velocity and $\mb{dl}_{n}$ an infinitesimal displacement. It was assumed that $\mb{\mathcal{A}}_2$ does not vary appreciably in the dimensions of each wave packet. The above AB phase accumulation leads to a phase difference 
\begin{equation}
	\frac{q}{\hbar}\oint \mb{dl}\cdot{\mathcal{A}}_2=\frac{q\Phi_B}{\hbar}
\end{equation}
between any two wave packets that propagate through opposite paths, where $\Phi_B$ is the magnetic flux enclosed by the interferometer, as expected \cite{aharonov59}.

Considering the electric version of the AB effect depicted in Fig 1(b), one conductor tube acquires a charge density $\rho_2(t)$ and the other $\rho_3(t)$ while the quantum particle is in a superposition of being inside each tube. The corresponding effective scalar potentials are $\mathcal{U}_2(t)$ and $\mathcal{U}_3(t)$, respectively, such that, according to Eq. (\ref{DE}), the total AB phase difference between the paths becomes
\begin{equation}
	-\int dt \frac{(\Delta E_{13}-\Delta E_{12})}{\hbar}=\frac{q}{\hbar}\int dt [\mathcal{U}_2(t)-\mathcal{U}_3(t)],
\end{equation}
as expected \cite{aharonov59}.

\section{Aharonov-Bohm effect with shielding} 

Previous works have used a quantized electromagnetic field to treat the magnetic \cite{santos99,choi04,pearl17b,li18,marletto19,saldanha19} and electric \cite{saldanha19} versions of the AB effect, with results similar to ours. The model presented so far provides a unified covariant treatment for these phenomena. Moreover, none of these previous works has considered the situation depicted in Fig. 1(c), where the long solenoid is coated with a perfect conductor shield. Most of the works that treat the AB effect considering an interaction of the classical field of the particle with the solenoid current or with the solenoid field do not consider the shielding situation either \cite{liebowitz65,boyer73,boyer02,peshkin81,vaidman12,kang13,pearl17a} (exceptions are Refs. \cite{kang15,saldanha16}). Note that the (classical) magnetic field produced by the passing quantum particle with a constant velocity $\mb{v}$ is given by $\mb{B}=\mb{v}\times\mb{E}/c^2$, where $\mb{E}$ is the (classical) electric field it produces \cite{griffiths}. So, if the perfect conductor shield cancels $\mb{E}$, it also cancels $\mb{B}$ inside the solenoid, since $\mb{B}$ is produced by the time variation of $\mb{E}$. Another way to understand the magnetic field cancellation inside the conductor shield is that in the particle rest frame only an electric field is produced, and the magnetic field appears with a Lorentz transformation acting in this field with the change of reference frame to one where the particle has a nonzero velocity. So, if the shield cancels the particle electric field in a region in the particle rest frame, the magnetic field in this region in the new frame should also be canceled. If the shield is a superconductor, due to the Meissner effect it completely cancels any magnetic field produced by an outside source in its interior, even a stationary field. The difficulty of explaining the AB effect in terms of fields in this situation with shielding is then obvious: if the shield prevents the field generated by the particle to penetrate inside the solenoid, how to explain the interaction of the particle field with the solenoid? Since experiments have confirmed the presence of the AB phase in this situation \cite{tonomura86} (with a toroidal magnet instead of a long solenoid, covered by a superconductor shield), any theory that aims to explain the origin of the AB phase must be capable of predicting this experimental result. In the following we show how our model can predict the presence of the AB phase in this situation, contrary to intuition. We shall see that this is possible precisely because we resort to a quantum model. 

In the scheme of Fig. 1(c), $\mb{J}_1$, $\mb{J}_2$ and $\mb{J}_3$  represent the current densities of the quantum particle, solenoid, and conductor shield, respectively. The interaction energy $\Delta E_{12}$ between the quantum particle and the solenoid results in the AB phase, as previously shown. The interaction energy $\Delta E_{13}$ between the quantum particle and the shield is not associated to the AB effect. It may affect the particle scattering, but in a way that is completely independent from the electromagnetic field inside the solenoid, so we will not consider it here. The term $\Delta E_{23}$ represents the interaction between the solenoid and the conductor shield mediated by the quantum electromagnetic field. Since the magnetic field generated by the shield precisely cancels the magnetic field generated by the quantum particle in the solenoid, we have $\nabla\times\mb{\mathcal{A}}_3(\mb{r}_2)=-\nabla\times\mb{\mathcal{A}}_1(\mb{r}_2)$ within the solenoid. So we may write $\mb{\mathcal{A}}_3(\mb{r}_2)=-\mb{\mathcal{A}}_1(\mb{r}_2)+\nabla F(\mb{r}_2)$ within the solenoid. Using Eq. (\ref{DE}), we  have
\begin{eqnarray}\label{Delta23}\nonumber
	\Delta E_{23}&=&-\int d^3r_2 \mb{J}_2(\mb{r}_2)\cdot\mb{\mathcal{A}}_3(\mb{r}_2)\\
	&=&-\int d^3r_2 \mb{J}_2\cdot[-\mb{\mathcal{A}}_1+\nabla F]=-\Delta E_{12},
\end{eqnarray}
where we used $\mb{J}_2\cdot (\nabla F)=\nabla\cdot(F\mb{J}_2)-F(\nabla\cdot \mb{J}_2)$, $\nabla\cdot\mb{J}_2=0$, and $\Delta E_{21}=\Delta E_{12}$.

So the total energy change of the vacuum state of the quantum electromagnetic field does not depend on the quantum particle path in the interferometer, since according to Eq. (\ref{Delta23}) $\Delta E_{23}$ cancels $\Delta E_{12}$. This means that the association of the AB phase difference to an entanglement between the particle path and the energy of the quantum electromagnetic vacuum state, as mentioned in  previous works \cite{marletto19,saldanha19}, does not seem to be adequate in this situation with shielding. Note that this is also true with a classical description of the electromagnetic field. When the solenoid is shielded, the magnetic energy $\int d^3r|\mb{B}_T(\mb{r})|^2/(2\mu_0)$, where $\mb{B}_T(\mb{r})$ represents the total magnetic field, does not depend on the particle path, since the particle magnetic field does not penetrate inside the solenoid and is not superposed to the solenoid field. 

But note that the energy $\Delta E_{23}$ can be associated to an exchange of  photons between the solenoid and the shield, not being related to a local interaction of the quantum particle with the electromagnetic field. {Considering that only local interactions of the quantum charged particle with the quantum electromagnetic field can affect the particle behavior, our \textit{locality assumption},  we predict the correct AB effect with a shielding of the solenoid.} Thus, only the exchange of photons between the charged particle and the solenoid, which results in the variation $\Delta E_{12}$ for the energy of the vacuum state of the quantum electromagnetic field, is to be considered (despite $\Delta E_{13}$, which is not related to the AB effect). So our local treatment correctly predicts that the AB effect should occur in the presence of the  perfect conductor shield around the solenoid, as observed in the experiments \cite{tonomura86}. In these experiments a superconductor shield is used, and we have the extra phenomenon of magnetic flux quantization. To apply our results to this case, we just have consider $\mb{J}_2$ as the total current density in the absence of the quantum particle (the solenoid current plus the current in the superconductor shield that accounts for the flux quantization) and $\mb{J}_3$ as the extra current density induced in the shield by the passage of the quantum particle. The conclusions are the same.

We can give further support for the \textit{locality assumption} used in the previous paragraph in another, simpler, scenario. Consider that we have a charge density $\rho_2$ in a region of space. When we approximate a charge $q_1$ (charge density $\rho_1$), it induces the appearance of another charge density $\rho_3$ in the system, for instance in a metallic object. With a quantum treatment for the electromagnetic field, we conclude that the perturbed energy of the electromagnetic vacuum state has a contribution $\Delta E_{12}+\Delta E_{13}+\Delta E_{23}$ with $\Delta E_{ij}$ given by Eq. (\ref{DE}) with null current densities. The Lorentz force law states that the charge $q_1$ must suffer an electrostatic force which is equal to $q_1$ times the local electric field. This force is minus the gradient of ($\Delta E_{12}+\Delta E_{13}$). Note that $\Delta E_{23}$ depends implicitly on the position of $q_1$, via $\rho_3$, such that its inclusion in the effective particle potential energy violates the Lorentz law. So the \textit{locality assumption} that only local interactions of the charge with the quantum electromagnetic field can affect the particle behavior lead us to a time evolution for the system which is compatible with the Lorentz force law, by disregarding $\Delta E_{23}$. So this assumption is essential to make the correct predictions in other situations with induced charges or currents, not only in the AB effect with shielding.

A discussion of the locality issues of the AB effect might be in order. In classical electrodynamics, the electromagnetic fields are seen as the responsible for the electromagnetic interactions. The electromagnetic potentials appear as mathematical tools used to simplify the calculations, since it is not possible to attribute definite values for these quantities due to their gauge-dependence. On this way, the description of the magnetic AB effect in terms of a local interaction of the quantum charge with the classical gauge-dependent potential vector may not be a completely satisfactory fundamental description of the problem. With the use of a quantum electromagnetic field, on the other hand, which is a more fundamental description, a clear physical picture of the problem emerges: the AB effect appears due to a photon exchange between the solenoid and the quantum particle, where both the solenoid and the particle interact locally with the quantum electromagnetic field. For this reason, we believe that the locality of the AB effect is more satisfactorily described with a quantum electromagnetic field.

\section{Aharonov-Casher effect}

 Our formalism also applies to the AC effect \cite{aharonov84}, represented in Fig. 1(d). The quantum particle is now a neutral particle with intrinsic magnetic moment $\bs{\mu}$, associated to a current density $\mb{J}_1=\nabla\times\mb{M}$ in the rest frame. When the particle moves with a nonrelativistic velocity $\mb{v}$, it acquires a charge density $\rho_1=\mb{v}\cdot\mb{J}_1/c^2$. The linear charge density $\lambda$ in the long wire produces an effective scalar potential $\mathcal{U}_2(\mb{r}_1)$ along the particle path. Using Eq. (\ref{DE}), we have
\begin{equation}
	\Delta E_{12}=\frac{1}{c^2}\int d^3r_1 \mb{v}\cdot \left[\mathcal{U}_2 \nabla\times \mb{M} \right]=\frac{1}{c^2}\mb{v}\cdot \left[ \mb{E}_2 \times \bs{\mu} \right],
\end{equation}
where $\mb{E}_2=-\nabla \mathcal{U}_2$ and we consider a negligible variation of this electric field in the dimensions of the particle wave function in each path. This expression leads to an AC phase difference  $\mu_0\mu\lambda/\hbar$ between the paths when the magnetic dipole is in the same direction as the wire axis, as expected \cite{aharonov84}. 

Following the suggestion of an anonymous referee, we can also consider the situation where the two paths indicated in Fig. 1(d) are in the interior of conductor tubes. On this way, the neutral particle always propagate in regions with a null electric field and the AC phase should disappear. In this case, electric charges are induced in the conductor tubes walls, and the energy associated to the quantum interaction of the  particle with these charges cancels the energy from its interaction with the charged wire. So a shield may cancel or not the AB or AC phase, depending on the specific system configuration.

\section{Conclusion}

We have presented a unified formalism to describe the electric and magnetic AB effects, as well as the AC effect, with the interactions being mediated by the quantum electromagnetic field, including situations with shielding. At a first sight, it could be imagined that the physical model of an exchange of photons between the quantum particle and the solenoid would not be able to describe the magnetic AB effect with a perfect conductor shield involving the solenoid, since the shield would block the photon exchange. But this is not the case. Charges and currents inside a perfect Faraday cage still interact with all charges outside it, exchanging photons with them. Although these interactions cannot produce force on the particles inside the cage, since their interaction with the shield cancels the effect of the external charges, they produce measurable results, the AB effect being an example.

\textit{Acknowledgements:} I greatly acknowledge Chiara Marletto and Vlatko Vedral for extensive and very useful discussions in this topic. This work was supported by the Brazilian agency CNPq. 


\begin{thebibliography}{31}%


\bibitem{ehrenberg49} W. Ehrenberg and R. E.  Siday, Proc. Phys. Soc. B \textbf{62}, 8 (1949).

\bibitem{aharonov59} Y. Aharonov and D. Bohm, Phys. Rev. \textbf{115}, 485 (1959).

\bibitem{liebowitz65} B. Liebowitz, Nuovo Cimento \textbf{38}, 932 (1965).

\bibitem{boyer73} T. H. Boyer, Phys. Rev. D \textbf{8}, 1679 (1973).

\bibitem{boyer02} T. H. Boyer, Found. Phys. \textbf{32}, 41 (2002).

\bibitem{peshkin81} M. Peshkin, Phys. Rep. \textbf{80}, 375 (1981).

\bibitem{vaidman12} L. Vaidman, Phys Rev. A \textbf{86}, 040101(R) (2012).

\bibitem{kang13} K. Kang, arXiv:1308.2093.

\bibitem{kang15} K. Kang, Phys. Rev. A \textbf{91}, 052116 (2015).

\bibitem{saldanha16} P. L. Saldanha, Braz. J. Phys. \textbf{46}, 316 (2016).

\bibitem{pearl17a} P. Pearle and A. Rizzi, Phys. Rev. A \textbf{95}, 052123 (2017).


\bibitem{santos99} E. Santos and I. Gonzalo, Europhys. Lett. \textbf{45}, 418 (1999).

\bibitem{choi04} M. Y. Choi and M. Lee, Curr. Appl. Phys. \textbf{4}, 267 (2004).

\bibitem{pearl17b} P. Pearle and A. Rizzi, Phys. Rev. A \textbf{95}, 052124 (2017).

\bibitem{li18} B. Li, D. W. Hewak, and Q. J. Wang, Found. Phys. \textbf{48}, 837 (2018).

\bibitem{marletto19} C. Marletto and V. Vedral, Phys. Rev. Lett. \textbf{125}, 040401 (2020).

\bibitem{saldanha19} P. L. Saldanha, 	Found. Phys. \textbf{51}, 6 (2021).


\bibitem{aharonov15} Y. Aharonov, E. Cohen and D. Rohrlich, Phys. Rev. A \textbf{92}, 026101 (2015).


\bibitem{tonomura86} A. Tonomura, N. Osakabe, T. Matsuda, T. Kawasaki, J. Endo, S. Yano, and H. Yamada, Phys. Rev. Lett. \textbf{56}, 792 (1986).


\bibitem{aharonov84} Y. Aharonov and A. Casher, Phys. Rev. Lett. \textbf{53}, 319 (1984).

	
\bibitem{cohen} C. Cohen-Tannoudji, J. Dupont-Roc, and G. Grynberg, Photons and Atoms (Wiley, New York, 1989).

\bibitem{mandel} L. Mandel and E. Wolf, Optical Coherence and Quantum Optics (Cambridge, New York, 1995).


\bibitem{cohen2} C. Cohen-Tannoudji, B. Diu, and F. Lalo\"e, Quantum Mechanics (Wiley, Paris, 1977), 2nd ed.

\bibitem{griffiths}
D. J. Griffiths, Introduction to Electrodynamics (Prentice Hall, Upper Saddle River, 1999), 3rd ed.


\end{thebibliography}

%

\end{document}